\documentclass[prl,twocolumn,superscriptaddress,amsfonts,amsmath,
floatfix]{revtex4}
\usepackage{graphicx}
\usepackage{psfrag}

\begin{document}

\title{Quantum stirring as a sensitive probe of 1D superfluidity}

\author{R. Citro}
\email{citro@sa.infn.it} \affiliation{Dipartimento di Fisica "E.R.
Caianiello", Universit{\`a} di Salerno, and  Unit{\`a} C.N.I.S.M.,
I-84081 Baronissi, Italy}
\author{A. Minguzzi}
\affiliation{Universit\'e Joseph Fourier, Laboratoire de Physique
et Mod\'elisation des Mileux Condens\'es, C.N.R.S. B.P. 166, 38042
Grenoble, France}
\author{F.~W.~J. Hekking}
\affiliation{Universit\'e Joseph Fourier, Laboratoire de Physique
et Mod\'elisation des Mileux Condens\'es, C.N.R.S. B.P. 166, 38042
Grenoble, France}

\begin{abstract}
We propose quantum stirring with a laser beam as a probe of
superfluid behavior
 for a strongly interacting one-dimensional Bose gas
confined to a ring. Within the Luttinger liquid theory framework,
we calculate the fraction of stirred particles per period as a
function of the stirring velocity, the interaction strength and
the coupling between the stirring beam and the bosons. The
fraction of stirred particles allows to probe superfluidity of the
system. We find that it crosses over at a critical velocity, lower
than the sound one, from a characteristic power law at high
velocities to a constant at low velocities. Some experimental
issues on quantum stirring in ring-trapped condensates are
discussed.
\end{abstract}

\maketitle

Progress in the ability to manipulate low-dimensional ultracold
atomic gases has stimulated the interest in fundamental properties
of one-dimensional (1D) Bose liquids
\cite{hellweg01_bec1d,goerlitz01_bec1d,richard03_bec1d}. A
Bose-Einstein condensate (BEC) of an atomic gas is known to
exhibit superfluidity. Experiments have confirmed the superfluid
behavior by demonstrating a critical velocity below which a laser
beam could be moved through the gas without causing
excitations\cite{stirring_raman_99,stirring_onofrio_00}, and an
irrotational flow through the creation of
vortices\cite{dalibard_stirring} and vortex
lattices\cite{abo_vortex} in both rotating and nonrotating traps.
For a Bose-Einstein condensate in a toroidal trap the observation
of a persistent flow has also been reported \cite{phillips_ring}.

Parametric pumping offers another way of inducing particle
transfer without creating excitations. In pumping, periodic (ac)
perturbations of the system yield a dc current. Indeed, this
current may be entirely adiabatic as long as the external
perturbations are slow enough such that the system always remains
in the instantaneous ground state. The number of particles
transferred in each cycle is then independent of the pumping
period $T$ and the integral of the current over a period is
quantized for a clean infinite periodic system with filled
bands\cite{thouless_qpt,altsh}.
Up to now, spectacular precision of quantization of the pumped
current has been achieved in experiments with nano-electronics
devices\cite{shilton_exp}.

Quantum pumping is intimately connected to quantum stirring.
Quantum stirring is accomplished by the cyclical variation of one
system parameter, while preserving the characteristic of a pump,
i.e. the orientation of the particle flow is fixed.  Quantum
stirring has been exploited to elucidate the nature of the
critical velocity in superfluid
liquids\cite{liquid_helium,stirring_raman_99,stirring_onofrio_00}.
We focus here on stirring a one-dimensional (1D) interacting Bose
gas with a laser beam in the regime where interaction effects are
especially strong, and we propose the fraction of stirred
particles as a measure of superfluid behavior. Although for a
homogeneous 1D Bose gas the superfluid fraction, defined as the
response to twisted boundary conditions and estimated from
ground-state quantities (see e.g. \cite{cazalilla_review}), is
always one independent of the interaction strength, it is a
relevant question to ask whether  the {\it out of equilibrium}
behavior of a 1D Bose gas under the effect of an external probe is
closer to the
 behavior expected for a superfluid (e.g. frictionless flow below a
 certain velocity threshold) or rather to a normal fluid (e.g. flow with drag).
The study of the stirred fraction gives a measure of the degree of
superfluidity of the fluid, i.e. a small stirred fraction
corresponds to superfluid-like and a unity stirred fraction
corresponds to normal-like behavior. This is complementary to the
onset of a drag force as manifestation of superfluid
behavior\cite{pavloff_02,astrakharchik_stirring}.

We consider $N$ bosons of mass $m$ confined onto a 1D ring of
circumference $L$, with contact interactions $v(x-x')=g
\delta(x-x')$ at zero temperature. The long-wavelength behavior of
this system at distances larger than the cutoff length
$\alpha=1/\rho_0=L/N$ is described by the Luttinger liquid
Hamiltonian in terms of the density and phase fluctuation modes of
the bosonic
field\cite{haldane_bose_81,giamarchi_book_1d,cazalilla_review}:
\begin{equation}
\label{eq:ham} H_0=\frac{\hbar}{2\pi}\int dx \lbrack \frac{v_s}{K}
(\nabla \phi(x))^2 + v_s K ( \nabla \theta(x))^2 \rbrack,
\end{equation}
 where the field $\phi(x)$ is related to the particle density  according to
 \begin{equation}\label{eq:density}
\rho(x)=\lbrack \rho_0-\frac 1 \pi \nabla \phi (x)\rbrack
\sum_{p=-\infty}^{\infty} e^{i2p(\pi \rho_0 x-\phi(x))},
\end{equation}
the field $\theta(x)$ corresponds to the phase of the superfluid,
and we have $[\phi(x),\nabla \theta(x')]=i \delta(x-x')$. In the
case of repulsive contact interaction between bosons, the
Luttinger parameters $v_s$ and $K$ used in (\ref{eq:ham}) are
obtained\cite{cazalilla_review} by: $v_s K = \frac{\pi\hbar
\rho_0}{m}$, as follows from galilean invariance, and
$\frac{v_s}{K} = \frac{g}{\hbar \pi}$ in the weak coupling limit,
while $\frac{v_s}{K} = \frac{\pi \rho_0}{\hbar
m}\left(1-\frac{8\rho_0 \hbar^2 }{m g} \right)$ in the strong
coupling limit. When the interaction goes to zero  $K$ goes to
infinity, while $K=1$ for infinitely strong hard-core interactions
(Tonks-Girardeau limit), where the problem can be solved by
mapping onto a gas of noninteracting fermions
\cite{girardeau_tonks_gas}. In this regime $2\pi \rho_0
\rightarrow 2 k_F$, with $k_F$ being the Fermi wavevector of the
corresponding mapped spinless fermions. The long-wavelength
properties of 1D dipolar gases are also described by
Eq.~(\ref{eq:ham}) with $K<1$~\cite{citro_dipolar}.


We describe next the effect of a barrier moving  with velocity $V$
through the fluid by introducing the time-dependent potential
$U(x,t)=U_0 \delta (x-Vt)$. In an experiment this could be
realized eg by stirring the gas by a blue-detuned laser beam.  The
Hamiltonian acquires an explicitly time-dependent term which
couples to the density:
\begin{equation}\label{eq:interaction}
  \delta H (t)=\int dx U(x,t) \rho(x).
\end{equation}
Using Eq.~(\ref{eq:density}) for the density and keeping only the
lowest, most relevant harmonics we may rewrite
(\ref{eq:interaction}) as
\begin{equation}\label{eq:interaction_bos}
 \delta H (t)= U_0\lbrack \rho_0-\frac 1 \pi \nabla \phi (Vt)+2\rho_0 \cos
(2\pi \rho_0 Vt -2
 \phi(Vt))\rbrack.
\end{equation}
The term proportional to $\nabla \phi$ is analogous to a slowly
varying chemical potential and can be absorbed in $H$ by a
redefinition of the field $\phi$, $\phi\rightarrow \phi-(K/v_s)
\int^{x} dx' U(x')$, while the last leading term in
Eq.~(\ref{eq:interaction_bos}) represents scattering of the bosons
off the
barrier with momentum close to $\pm 2\pi \rho_0$.
In the Tonks Giradeau\cite{girardeau_tonks_gas} limit it describes
the backscattering of right-movers into left-movers, i.e.
processes with momentum close to $\pm 2k_F$.  During its motion
the barrier drags along a part of the bosons.
We are interested in the {\em stirred fraction} $N_\mathrm{stir}/N$
 i.e. the  fraction of particles transported per period $T=L/V$ by the moving
barrier, and related to the particle current as $N_\mathrm{stir}=
\frac{1}{2\pi}\int_0^T dt I(t)$. If the barrier height is
infinitely large, the fraction of stirred particles per period is
quantized, i.e. $N_\mathrm{stir}/N=1$, independently of the
interaction strength.
If the barrier height is finite, the stirred fraction is in
general smaller than one and we  show that it is related to the
degree of correlations in the system. We analyze perturbatively
the regimes of weak and large barrier for arbitrary interaction
strength and treat exactly the Tonks-Girardeau regime.

{\it Weak barrier}- In the weak barrier limit we perform a
perturbative analysis of the current generated by the stirring
Hamiltonian $\delta H$. As customary in Luttinger liquid formalism
we introduce the particle density of right(left) movers related to
the fields $\theta(x)$ and $\phi(x)$ as $\rho_{R(L)}\simeq
\frac{\rho_0}{2}\pm\lbrack \nabla \theta (x) \mp \nabla \phi(x)
\rbrack$. The particle current at low energy is $J(x)\sim
\nabla\theta(x)$; since  it involves the difference in the number
of right and left movers, the term proportional to $\nabla \phi$
in Eq.~(\ref{eq:interaction_bos}), which does not distinguish
between left and right movers, plays no role in generating the
particle current. On the contrary, the third, backscattering term
in Eq.~(\ref{eq:interaction_bos}), can lead to the generation of a
current which we define of {\it backscattering}, $I_b$. In fact,
addition of  the moving-barrier potential breaks the continuous
chiral symmetry\cite{chamon_long} violating the conservation of
the axial charge $N_R-N_L$, where $N_{R(L)}=\int dx \rho_{R
(L)}(x)$. In the lowest order perturbation theory the
backscattering current is given by
$I_b^0=\frac{i}{\hbar}[N_L,\delta H]=-\frac{i}{\hbar}[N_R,\delta
H]$. In our specific case, by using the bosonized expression of
the density operators and of the stirring Hamiltonian, the
resulting backscattering-current operator is $I_b^0=i \gamma(t)
\tilde{n}(t)-\text{h.c.}$, where $\gamma(t)=U_0 e^{i 2\pi \rho_0
Vt}$ and $\tilde{n}\sim \rho_0 e^{i 2\phi(Vt)}$, and it is
characterized by the backscattering frequency $\omega_b=2 \pi
\rho_0 V$. Linear response theory yields the backscattering
current to second order in the barrier strength $U_0$ as
\begin{eqnarray}\label{eq:curr}
I_b&&\approx i\int_{-\infty}^t dt' \langle [I_b^0(t),\delta H(t')]
\rangle_{H_0}
\end{eqnarray}
and turns out to be related to the Fourier transform of the
Green's function of the  backscattering operator $e^{i 2\phi(Vt)}$
 at the characteristic frequency $\omega_b$.
In the thermodynamic limit $N,L\to \infty$ with $\rho_0 = N/L$
constant and for small stirring velocity, the resulting
backscattering current is given by
\begin{equation}
\label{eq:curre_result} I_b\approx\frac{(2
\pi)^{2K-1}}{\Gamma(2K)}\frac{U_0^2}{(\hbar v_s)^2}
\left(\frac{V}{v_s}\right)^{2K-2}  2\pi \rho_0 V,
\end{equation}
with $\Gamma$ being the Euler Gamma function. The fraction of
stirred particles  is readily obtained from the backscattering
current according to $N_\mathrm{stir}/N =I_b /\omega_{b}$. In the
Tonks-Girardeau limit $K\rightarrow 1$, Eq.(\ref{eq:curre_result})
yields $N_\mathrm{stir}/N\propto (U_0/\hbar v_s)^2$, i.e. the
result is independent of the frequency $\omega_b$ and hence
adiabatic~\cite{thouless_qpt}. In the small $\omega_b$ limit this
result is in agreement with the exact  calculation of the fraction
of stirred particles, as shown below. Note that  as the Luttinger
liquid theory is an effective low-energy model, it describes
correctly the system at frequencies $\omega_b < 2 \pi v_s/\alpha$,
hence the expression (\ref{eq:curre_result}) is valid only if
$V<v_s$, and cannot treat the supersonic regime. By recalling that
the power-law dependence in Eq.~(\ref{eq:curre_result}) originates
from the excitation of sound waves in the quasi-one-dimensional
geometry, we can also determine the smallest velocity for which
Eq.~(\ref{eq:curre_result}) holds in  the case of a ring of finite
length. In this case no excitations are possible below the lowest
velocity $V_{low}=v_s/N\sim \pi\hbar /m L$ corresponding to the
momentum of the lowest bosonic mode on the ring. The value of
$V_{low}$ found agrees with the one obtained by using a
Gross-Pitaevskii approach for $K\gg 1$ \cite{buchler}. Thus as a
main result we find that at the critical velocity $V_{low}$ the
fraction of stirred particles crosses from a power-law to a
constant (adiabatic regime). Note also  that the adiabatically
stirred fraction decreases with decreasing interaction strength as
$1/\Gamma(2 K)$: when $K$ grows, the system becomes more
superfluid, hence the interaction with the external barrier
decreases and $N_\mathrm{stir}/N\rightarrow 0$.

\begin{figure}
\includegraphics[width=85mm,height=56mm,angle=0]{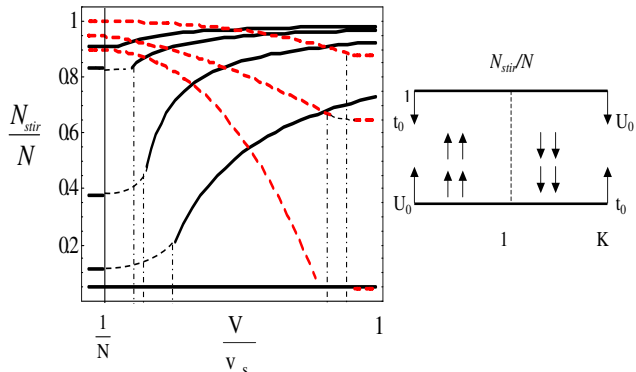}
\vspace{-5mm} \caption{(Color online) Left panel: Fraction of
stirred particles as a function of the stirring velocity (in units
of $v_s$), obtained matching Eqs.(\ref{eq:curre_result}) and
(\ref{eq:curre_result_1}) at $V_{U_0}^\star=
V_{t_0}^\star=V^\star$ through the RG considerations (see text).
(Black) Continuous line: $N_\mathrm{stir}/N$ at decreasing barrier
strengths $U_0/(\hbar v_s)$ from 1. (upper curve) to 0.2 (lower
curve) for $K>1$, fixed as 2; (red) dashed line:
$N_\mathrm{stir}/N$ at increasing tunneling strength $t_0/(\rho_0
\hbar v_s)$ from 0.2 (upper curve) to 0.6 (lower curve) for $K<1$,
fixed at 0.5. The vertical dashed-dotted lines indicate
$V^\star/v_s$. The vertical straight line indicates the critical
velocity $V_{low}$ below which the stirring is adiabatic. Right
panel: Summary of the RG flow for the barrier potential $U_0$ and
tunneling strength $t_0$ (left and right edge arrows) and
$N_\mathrm{stir}/N$ (arrows in the middle of the frame) at varying
the interaction $K$ approaching the adiabatic regime. The stirred
fraction is the analogous for neutral particles of the conductance
for a 1D electron gas with a barrier \cite{kane_fisher_prl}}
\label{fig1}
\end{figure}

The results obtained above are consistent with a treatment based
on the perturbative renormalization group (RG)
approach~\cite{kane_fisher_prl}. In this approach the scaling of
the potential $U_0$ with frequency $\omega$ is obtained from the
flow equation $dU_0/dz = (K-1)U_0$ where $dz = d\omega/\omega$. As
a function of $K$, two regimes are distinguished. When $K>1$ the
barrier is irrelevant: $U_0$ decreases as $\omega$ is decreased
from $v_s/\alpha$ down to $\omega_b\sim V/\alpha$. For an infinite
system, $U_0$ and hence $N_\mathrm{stir}/N$ scale to zero as
$\omega_b \to 0$; for a finite system the RG procedure should
stopped when  $\omega_b \sim v_s/L$, i.e. for $V \sim v_s/N$; this
is the regime where we find a residual adiabatically stirred
fraction, independent of $V$. For $K<1$, e.g. in the dipolar gas,
$U_0$ and hence $N_\mathrm{stir}/N$ grow under RG, i.e. the
barrier is a relevant perturbation. This is shown in the right
panel of Fig.\ref{fig1}. Perturbation theory breaks down when
$N_\mathrm{stir}/N \sim 1$, i.e. at the velocity $V_{U_0}^\star=
v_s (U_0/\hbar v_s)^\frac{1}{1-K}$ and the RG flow must be
stopped. The behavior beyond this breakdown point is described by
an effective weak-link tunneling model\cite{kane_fisher_prl}.

{\it Weak link limit}-
The large-barrier limit is equivalent to a ring cut
at the position of the delta barrier, and we treat
the residual tunneling $t_0$ between the two ends of the ring as a
perturbation
\cite{kane_fisher_prl}. In this case the bosonized Hamiltonian
corresponding to the hopping across the weak-link can be obtained
by a duality transformations\cite{kane_fisher_prl}, $\phi
\rightarrow \theta$ and is given by $\delta H \sim t_0 \cos( 2\pi
\theta(Vt))$. Its contribution to the particle current ({\it
tunneling} current $I_t$) can be calculated in the linear response
regime and its explicit expression for an infinitely long ring is:
\begin{equation}\label{eq:curre_result_1}
I_t=\frac{(2\pi)^{\frac 2 K -1} t_0^2}{(\rho_0 \hbar v_s)^2
\Gamma(2/K)}\left(\frac{V}{v_s}\right)^{\frac 2 K-2} 2 \pi \rho_0
V.
\end{equation}
In the presence of tunneling the stirred fraction of particles is
$N_{stir}/N=1-I_t/\omega_b$, where $I_t/\omega_b$ is the fraction
of tunnelled particles, not stirred. In the hard-core limit
($K=1$) the stirred current will be again linear in the frequency
of the stirring. We thus recover the adiabatic limit.
Under the RG flow,
the tunneling becomes relevant for interacting bosons with contact
repulsion ($K>1$)
therefore, upon decreasing the stirring velocity the effective
tunneling strength increases, thereby decreasing the stirred
particle fraction, again shown in the right panel of
Fig.\ref{fig1}. Perturbation theory breaks down when the
effective tunneling strength reaches unity and the RG flow must be
stopped at $V_{t_0}^\star=
v_s (t_0/\rho_0\hbar v_s)^\frac{K}{K-1}$ , then
the stirred fraction of particles is governed by the previous weak
barrier limit. The results for the dependence of the stirred
fraction of particles on the velocity $V$ is shown in
Fig.~\ref{fig1}. The results explicitly show  a difference
in the regime with $K>1$ (short-range interactions) and $K<1$
(dipolar interactions).
In the latter case the stirred fraction decreases at increasing
velocities, where the system tends towards superfluid behavior.
Since a dipolar gas is characterized by a quasi-crystal order
phase at increasing density\cite{citro_dipolar}, the result can be
interpreted as an inefficiency of the stirring in creating an
excitation in the ordered state.

\begin{figure}[tb]
\hspace{5mm}
\includegraphics[width=60mm,height=45mm,angle=0]{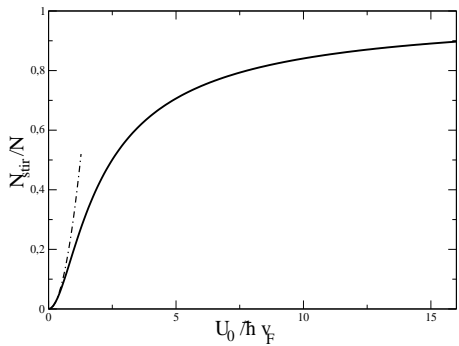}
\vspace{5mm}
 \caption{Fraction of stirred particles for a Tonks-Girardeau gas (i.e. case
 $K=1$) as a function of the barrier strength $U_0/(\hbar v_s)$,
  in the adiabatic limit from Thouless expression
  Eq.(\ref{eq:charge_final}) (solid line) and analytical small-$U_0$
  behavior (dot-dashed line) as explained in the text.}
\label{fig2}
\end{figure}

{\it Non-perturbative analysis}- In the Tonks-Giradeau limit
($K=1$) a time-dependent Fermi-Bose (FB) mapping
\cite{girardeau_tonks_gas} is employed to generate exact solutions
of the problem\cite{girardeau_delta} and the current is calculated
exactly. The time dependent version of the FB mapping permits to
write the exact many-body wavefunction  of $N$ impenetrable bosons
on a ring as
$\Psi_B(x_1,\ldots,x_N;t)=A(x_1,\ldots,x_N)\Psi_F(x_1,\ldots,x_N;t)$,
where $A$ is a unit antisymmetric function
$A(x_1,\ldots,x_N)=\prod_{1\le j<k\le N} \text{sgn}(x_k-x_j)$, and
$\Psi_F(x_1,\ldots,x_N;t)=C \det_{i,j=1}^{N} \psi_i(x_j,t)$ is the
wave function for any ideal Fermi gas, $\psi_i(x_j,t)$ being the
solutions of the one-body time-dependent Schroedinger equation in
the external potential $U(x,t)$. Starting from the above many-body
wavefunction, we evaluate the Tonks-Girardeau particle current
density in terms of the one-body density matrix $\rho_1(x,y)=\int
dx_2...dx_N \Psi_B^*(x,\ldots,x_N;t) \Psi_B(y,\ldots,x_N;t)$ as
$J(x)=-(\hbar/2m i) [\partial_{r}\rho_1(x+r/2,x-r/2)]_{r=0}$.
Although $\rho_1(x,y)$ for a TG gas is very different from the one
of a Fermi gas due to the presence of the mapping function $A$, we
find that the latter has no effect on the current, which then
coincides with the current of an ideal Fermi gas. In the adiabatic
limit  $V\le \pi \hbar/mL$ the particle current and the stirred
fraction produced by the slow variation of the stirring potential
can then be evaluated by following the adiabatic expansion of
Thouless for an ideal Fermi gas\cite{thouless_qpt}, i.e.
\begin{equation}\label{eq:charge_thouless}
N_s=\frac{i}{2\pi}\frac {\hbar^2}{m L}\int_0^\tau d
t\sum_{\ell,j\ne
  0}\frac{f_\ell(1-f_j)}{(\epsilon_\ell-\epsilon_j)}\lbrack \langle
\psi_j|\dot{\psi_\ell}\rangle \langle \partial_x \psi_\ell
|\psi_j\rangle
  +h.c.\rbrack,
\end{equation}
where $f_{\ell,j}$ is the fermionic  probability occupation
function of the state $\ell,j$. In the case of a blue-detuned
laser field  piercing the ring at a position $x=0$ and modelled by
the potential $U_0 \delta(x)$, the appropriate orbitals
$\psi_i(x_j,t)$ are  the $L$-periodic free-particle energy
eigenstates satisfying at $x=0$ the cusp condition. The complete
orthonormal set of even parity $\psi_n^{(+)}$ and odd parity
$\psi_n^{(-)}$ eigenstates are $\psi_n^{(+)}(x)=(e^{i k_n x}+e^{-i
k_n (x-L)})/{\cal N}_n$ and $\psi_n^{(-)}(x)=\sqrt{\frac 2 L}
\sin(2n\pi x /L)$,
where $k_n$ are obtained from the trascendental equation $k_n \tan
(k_nL/2)=mU_0/\hbar^2$ (for $U_0 \to \infty $  we have $k_n=\pi
(2n+1)/L$, in agreement with  \cite{girardeau_tonks_gas})
 and ${\cal N}_n=\sqrt{2L [1+\sin(k_nL)/k_nL)]}$,  with  $n$  running from
1 to $\infty$. The $N$-fermion ground state is obtained by
inserting the lowest N orbitals into the determinant above (Fermi
sea),
and  using the exact orbitals $\psi_n^{(\pm)}$ as instantaneous
ground state we obtain from (\ref{eq:charge_thouless})
\begin{equation}\label{eq:charge_final}
N_s=64 \sum_{\ell,j} (f_\ell-f_j)
\frac{\sin^2(k_jL/2)}{1+\sin(k_jL)/k_jL} \frac{(k_jL)^2
4\pi^2\ell^2}{[(k_jL)^2 -4\pi^2\ell^2 ]^3}
 \end{equation}
For a weak barrier  by using the small-$U_0$ expression for $k_n$
we obtain $N_s/N \simeq 0.32  (U_0/\hbar v_s)^2$, which scales as
the $K=1$ limit of the backscattered current in
Eq.(\ref{eq:curre_result}) because $v_s=\hbar k_F/m$ for $K=1$
\cite{note2}. For an infinitely strong barrier using the $U_0\to
\infty $ limit of $k_n$ it's
straightforward to verify that
 the particle transport is
quantized\cite{thouless_qpt}, i.e. all the particles are dragged
by the barrier and $N_\mathrm{stir}/N=1$. This is shown in
Fig.\ref{fig2}, where the stirred faction of particles is plotted
as a function of the barrier strength.

{\it Experimental issues on condensates in closed loop waveguide}
-A possible way of achieving experimentally an annular condensate
with strong transverse confinement is to use a magnetic toroidal
trap, as reported in \cite{gupta_05,arnold_06,phillips_ring}.
Experimentally the stirring of hydrodynamic flow in a BEC by a
blue-detuned laser beam, has been analyzed by calorimetric
method\cite{stirring_raman_99} and phase contrast
imaging\cite{stirring_onofrio_00}.
The onset of a drag force has been shown by the asymmetry in the
density profile, defined as the difference between the peak column
density in front and behind the laser beam, as a function of the
stirring velocity above a critical velocity. The space integral of
the density asymmetry is analogous to the fraction of stirred
particles calculated above.
Recently, the persistent flow of Bose-condensed atoms in a
toroidal trap has also been observed\cite{phillips_ring}.
A variant to such experiment by the addition of a cyclic moving
plug beam could be a valuable realization of the present proposal.

In conclusion, superfluid flow in a ring geometry raises
interesting new possibilities. With the use  of a
moving barrier acting as a quantum stirrer, the analog of
quantization of particle transport for electron systems could be
realized for a gas of atoms as an alternative probe of
superfluidity.

{\it Acknowledgments}- The authors would like to thank L. Glazman,
M. Girardeau, E. Orignac and E. Wright for useful suggestions.
This work was financially
supported by the European community as a part of a Marie Curie
Program and by the MIDAS STREP project.


\begin{thebibliography}{10}
\bibitem{note2} The prefactor of $N_{stir}/N$ obtained from Eq.(\ref{eq:curre_result}) is non-universal, as it depends on the short-distance cutoff $\alpha$, here set equal to $1/\rho_0$.
\end{thebibliography}

\end{document}